# Electrical addressing of exceptional points in compact plasmonic structures

Hoon Yeub Jeong[1], Yeonsoo Lim[1], Jungho Han[1], Soo-Chan An[1], and Young Chul Jun[1,2,*]

[1]Department of Materials Science and Engineering, Ulsan National Institute of Science and Technology (UNIST), Ulsan 44919, Republic of Korea

[2]Graduate School of Semiconductor Materials and Devices Engineering, UNIST, Ulsan 44919, Republic of Korea

Email: [*]ycjun@unist.ac.kr

(Dated: April 3, 2023)

**Abstract**

Exceptional points (EPs) are degenerate singularities in a non-Hermitian system that can be induced by controlling the interaction between resonant photonic modes. EPs can enable unusual optical phenomena and significantly enhance the optical sensitivity under small perturbations. However, most studies thus far have been limited to static photonic structures. In this study, we propose and experimentally demonstrate electrically addressable EP in a plasmonic structure. Inspired by optical microcavity studies, we employ a localized spoof plasmon structure that supports circulating plasmonic modes in a compact single-resonator geometry. The plasmonic modes are perturbed by an angled metal line, and the interaction between the plasmonic modes is electrically controlled using a varactor. Continuous electrical tuning of the varactor capacitance facilitates simultaneous coalescence of the real and imaginary parts of the eigenfrequency, allowing the direct addressing of EPs. We first investigate the eigenmodes and their coupling in localized plasmonic structures using numerical simulations. We then present experimentally measured spectra that manifest the coalescence of the two resonant modes in both the resonance frequency and linewidth. Electrically addressable EPs in compact plasmonic structures may provide exciting opportunities for highly functional and tunable elements in integrated device platforms.

# 1. Introduction

Exceptional points (EPs) are degenerate singularities in a non-Hermitian (or non-conservative) system that exchange energy with its surrounding environment [1, 2]. An EP can be induced by coupled optical resonances in a non-Hermitian system when two or more eigenvalues and their corresponding eigenvectors coalesce simultaneously in parameter space. EPs have been studied for novel laser and nonlinear optical systems, ultrasensitive measurements, asymmetric mode switching, and exotic scattering and topological structures [1–6]. EPs and non-Hermitian photonic systems have attracted considerable attention and may provide new platforms for photonic device design.

Optical microdisks or microtoroidal cavities have been used as a general platform for non-Hermitian photonics [7]. For example, EPs can be realized in an optical microdisk (or microtoroid) loaded with two particles on its circumference [8, 9]. These particles perturb the circulating optical modes in the cavity and induce optical coupling between the resonant modes. By controlling the position of the particles on the circumference, two optical modes can be gradually tuned to exhibit crossing or anti-crossing in the real and imaginary parts of the complex eigenfrequency. Based on these particle-loaded cavities, chiral lasing [10], perfect absorption [11], and enhanced optical sensing [12, 13] have been realized using EPs.

Although EPs have attracted considerable attention, most studies have been limited to static structures. In such static photonic structures, it is often difficult to directly address the exact EP because both the real and imaginary parts of the eigenfrequency must coalesce simultaneously. In this regard, dynamically and continuously tunable systems in a compact, integrated device platform have unique merits for the realization and practical application of EPs. Recently, electrically tunable EPs have been demonstrated in the terahertz region using ionic electrolyte gating of graphene [14]. The interaction between light and the vibrational modes of organic molecules was exploited to electrically control the reflection near the EP. Switching of the direction of reflectionless light propagation at exceptional points was also studied using phase-change materials [15]. Generally, dynamically tunable photonic systems are important for many practical applications [16-19].

Here, we present a different approach based on a compact single-resonator structure. Inspired by optical microcavity studies, we employ a spoof plasmon structure that supports circulating plasmonic modes. The plasmonic modes are perturbed by an angled metal line, and an EP is realized in a compact resonator geometry. In our design, a voltage-controlled capacitor (varactor) is introduced to electrically address the EP.

Spoof surface plasmons in corrugated metal surfaces can allow tight field confinement even at low frequencies (such as the microwave region), similar to surface plasmon polaritons (SPPs) at optical frequencies [20–25]. At low frequencies, electromagnetic fields cannot penetrate metals and tight field confinement via plasmonic responses cannot be ordinarily achieved. However, subwavelength corrugations on a metal surface can induce plasmonic responses even at low frequencies by playing a role similar to that of field penetration into metals at optical frequencies [26, 27]. The electromagnetic responses of spoof plasmon waves can be controlled using corrugation geometry, which provides a flexible platform for device design and response control. Spoof plasmon structures can also be applied to localized resonator geometries by circularly bending one-dimensional grooves, which are called localized spoof plasmons (LSPs) [28–37]. The resonance frequency and multipole mode excitation can be controlled by adjusting the structural parameters.

In this work, we control coupled plasmonic resonances using a varactor and an angled metal line to realize an EP in a compact plasmonic structure. The response of coupled plasmonic modes can be dynamically and continuously tuned by applying a bias voltage and inducing the capacitance variance of the varactor [38–40]. We first investigate the eigenmodes and their coupling in the LSP structures using numerical simulations. Then, we present the experimental results and discuss the measured features of the EP. The measured spectra manifest the coalescence of the two resonant modes in both the resonance frequency and linewidth simultaneously. Electrically addressable EPs in our ultrathin, printed plasmonic structures may facilitate highly functional and tunable elements in integrated device platforms.

## 2. Results and Discussion

Figure 1 shows a schematic of our spoof plasmon resonator. It consists of a two-layer printed circuit board (PCB) structure whose front side (copper plane) is patterned into an LSP structure, while the back side has a metal (copper) line (Figure 1A). This metal line works as a microstrip line in our sample. The total radius ($r_1$) of the LSP structure is 18 mm, including the outer grooves, while the radius ($r_2$) of the inner circle completely filled with metal is 10 mm (Figure 1B). The LSP structure is surrounded by another outer metal plane. The gap between the LSP structure and outer metal plane was 1 mm (see the Methods section). Subsequently, this gap was connected by a varactor to electrically control the response of the plasmonic resonator.

The LSP structure supports multipole resonances owing to spoof plasmon waves circulating along the corrugated circumference [41–43]. These multipole resonances in the LSP structure are excited when the plasmon waves approximately satisfy $n\lambda_{sp} \approx 2\pi r_1$, where $\lambda_{sp}$ is the wavelength of the spoof plasmons, $2\pi r_1$ is the circumference of the structure, and a positive integer $n$ is the azimuthal mode number. In our design, we use an angled metal line on the back side to perturb the plasmonic modes in the LSP resonator and control the interaction between them. The metal line has an angle, θ, between the two metal segments (Figure 1A). This design is inspired by an optical microdisk (or toroidal resonator) loaded with two particles on its circumference, which has been used to induce EPs at optical frequencies [8–12].

To understand how the capacitance of the varactor affects the spoof plasmon resonances, we first conducted COMSOL eigenfrequency/eigenmode simulations (see Methods) assuming a simplified geometry without a bottom metal line (Figure 2); i.e., we only considered the LSP structure loaded with a varactor on the front side. Initially, at zero capacitance, two degenerate modes exist for each resonance. As the capacitance increases, one mode gradually redshifts (dashed line), while the other mode remains unaffected (solid line) (Figure 2B). The eigenmode profiles in Figure 2C show that when the varactor is located at the node (or zero field) positions, the eigenmode remains unaffected. In contrast, when the varactor is located at non-zero field positions, a gradual redshift occurs with increasing capacitance. This redshift can be explained by the *LC* resonator model: $f \sim 1/\sqrt{LC}$. The circular shape of the LSP structure generates inductance *L* while the gap between the LSP and the outer metal plane induces capacitance *C*. Therefore, the resonance frequency *f* gradually decreases (i.e., redshift) with increasing capacitance.

We now consider the perturbation of the plasmonic modes by the metal line (Figure 3). First, we consider the case in which the length of the two metal segments ($L_{pert}$) gradually increases until they touch ($L_{pert}$ = 19 mm) and form a continuous metal line while maintaining θ = 0° (i.e., straight line) (Figure 3A). Figure 3B shows how the eigenfrequency changes with increasing $L_{pert}$. At zero-segment length, two degenerate modes exist for each resonance. The resonant frequency of one mode gradually redshifts as the length of the segment increases, whereas the resonant frequency of the other mode remains almost constant.

Figure 3C shows the field profiles ($E_z$) of the eigenmodes. Here, we assumed that the two segments touched each other ($L_{pert}$ = 19 mm, θ = 0°). A comparison of the mode profiles with the resonance frequencies in Figure 2B shows that when the metal line is located at the node (or zero field) position of the eigenmode (such as 4.75 GHz and 5.58 GHz), the resonance frequency remains nearly constant regardless of the segment length $L_{pert}$. However, when the metal line lies at non-zero field positions, the eigenmode frequency gradually shifts with the segment length.

Figure 4A shows a photograph of the real varactor-loaded sample. The varactor was placed in the gap between the spoof LSP structure and surrounding metal plane on the front side, and soldered to them. The capacitance of the varactor can be electrically tuned from 2.22 to 0.3 pF by applying a DC voltage bias from 0 to 20 V (i.e., the capacitance decreases as the voltage increases). Owing to the patterning of the front side into a spoof plasmon structure, it was possible to position the varactor in the gap. Copper wires were soldered at the center of the spoof plasmon structure and at the corner of the ground plane to apply the bias voltage across the gap (see Figure 4A). Because most plasmonic fields are confined near the gap and wings of the LSP structure, the soldering of copper wires at the center of the spoof plasmon structure and at the corner of the surrounding metal plane have minimal impact on the spoof plasmon resonances.

The LSP resonance can also be tuned using the metal-line angle. Figures 4B and 4C show the results of the eigenmode simulations in the real sample including capacitances of 2.22 pF and 0.3 pF, respectively. The angle of the metal line on the back side is gradually varied from θ = 0° to 30°. Two eigenmodes exist in the frequency range of 5 ~ 5.5 GHz, and the resonance frequency of one of these modes can be significantly adjusted by the capacitance. Figures 4B and 4C indicate that the metal-line angle can be used as a parameter to control the eigenmode frequency and the interaction between the two modes.

The time-domain simulation in Supplementary Video V1 directly demonstrates the role of the metal line. For simplicity, we considered a one-dimensional (straight) spoof plasmon waveguide (see also Figure S1 in Supplementary Material). The metal line on the backside was aligned to pass through the middle of the plasmon waveguide. A point dipole source was used to excite the propagating waves along the metal line. Video V1 shows that the metal line splits the incident waves into the two opposite directions of the spoof plasmon waveguide. In the case of our circular LSP resonator, the split waves become clockwise and counterclockwise circulating plasmon waves. Their interactions form different resonant modes depending on the metal-line angle.

Comparing the dashed circles in Figures 4B and 4C, we notice that the two eigenmodes have interactions between 15° and 18°. Because an EP can result from the interaction of two resonant modes, this indicates that an EP may exist in this range of parameters (metal-line angle and capacitance). Performing additional eigenfrequency simulations, a three-dimensional, complex eigenfrequency surface was obtained and visualized in parameter space (Figures 5A and 5B). The real and imaginary parts of the complex eigenfrequency correspond to the frequency (Re[$f$]) and damping rate (Im[$f$]) of the resonant mode, respectively. Then, the linewidth (full width at half maximum, FWHM) in the resonance spectrum becomes 2Im[$f$]. Figure 5A shows the real part of the complex eigenfrequency as a function of the angle and capacitance, whereas Figures 5B shows the imaginary part of the complex eigenfrequency. These demonstrate that the two eigenmodes interact and produce a nontrivial topology in parameter space.

More specifically, Figures 5C and 5D show the real and imaginary parts of the complex eigenfrequency at θ = 15.83°, respectively. The two eigenmodes nearly coalesce around 5.32 GHz and 0.585 pF; i.e., the complex eigenfrequencies of the two modes overlap in both the real and imaginary parts, indicating the occurrence of an EP. Below 0.585 pF (i.e., below the EP), the real parts of the two complex frequencies are very close, whereas the imaginary parts are largely separated (i.e., the two modes have different linewidths). Above 0.585 pF (i.e., above the EP), the situation is reversed; the real parts are largely separated into higher- and lower-frequency modes, whereas the imaginary parts are similar (i.e., the two modes have similar linewidths). Figure S2 shows additional complex frequency data for different metal-line angles (from θ = 15° to 17.8°). Between θ = 15° and 16°, the two eigenmodes are very close for both the real and imaginary parts. The two modes nearly coincide at θ = 15.83° and manifest the features of EPs.

Figure S3 in Supplementary Material shows the field profiles ($E_z$) of the simulated eigenmodes (θ = 15.83°) at 2 pF (i.e., away from the EP). The two eigenmodes at 5.27 GHz and 5.34 GHz are significantly different (particularly in the upper region of the LSP structure). However, near the EP condition, the field profiles of the two eigenmodes become similar (Figure S4), as expected for coalesced eigenmodes.

Generally, an EP occurring from two interacting modes can be described using the following non-Hermitian Hamiltonian [1, 44]:

$$\widehat{H} = \begin{pmatrix} \omega_1 + i\gamma_1 & \kappa \\ \kappa & \omega_2 + i\gamma_2 \end{pmatrix} = \begin{pmatrix} \omega_0 + i\gamma_0 & 0 \\ 0 & \omega_0 + i\gamma_0 \end{pmatrix} + \begin{pmatrix} \Delta\omega + i\Delta\gamma & \kappa \\ \kappa & -\Delta\omega - i\Delta\gamma \end{pmatrix}, \quad (1)$$

where $\omega_i$ and $\gamma_i$ are the resonant angular frequency and damping rate, respectively, for each mode (i = 1, 2), and $\omega_0 = \frac{\omega_1+\omega_2}{2}$, $\gamma_0 = \frac{\gamma_1+\gamma_2}{2}$, $\Delta\omega = \frac{\omega_1-\omega_2}{2}$, and $\Delta\gamma = \frac{\gamma_1-\gamma_2}{2}$. $\kappa$ is the complex coupling constant between the two modes. Then, the complex eigenfrequencies $\omega_\pm$ of this Hamiltonian are given by

$$\omega_\pm = \omega_0 + i\gamma_0 \pm \sqrt{\kappa^2 + (\Delta\omega + i\Delta\gamma)^2}. \quad (2)$$

And the corresponding eigenvector becomes

$$\boldsymbol{a}_\pm = \begin{pmatrix} 1 \\ \frac{-\Delta\omega - i\Delta\gamma \pm \sqrt{\kappa^2 + (\Delta\omega + i\Delta\gamma)^2}}{\kappa} \end{pmatrix}, \quad (3)$$

which satisfies $\widehat{H}\boldsymbol{a}_\pm = \omega_\pm \boldsymbol{a}_\pm$. Both the eigenfrequency and eigenvector have the same form of the square root term in Eq. (2) and (3). Therefore, when the square root term becomes zero, the two eigenvalues $\omega_\pm$ and the corresponding eigenvectors $\boldsymbol{a}_\pm$ coalesce simultaneously, and an EP occurs. Our eigenmode simulations show the simultaneous coincidence of the real and imaginary parts of the complex eigenfrequencies, indicating the occurrence of an EP in our spoof plasmon structure.

We also conducted more eigenfrequency simulations and confirmed the characteristic square-root dependence of the mode splitting near an EP. In the simulation, we gradually increased the refractive index of the surrounding medium (n = 1 + Δn) and measured changes in the mode splitting (Figure S5). Fitting of this mode splitting in the log-log plot perfectly matches a square-root curve (i.e., slope = 1/2) under small perturbations. This is clearly different from other mode-splitting mechanisms (diabolic points) which exhibit linear dependence under perturbations [12, 13].

We now present our experimental results. We prepared five samples with different metal-line angles of θ = 14°–18° in increments of 1°. For each sample, we gradually varied the bias voltage of the varactor from 0 to 20 V and measured the reflection spectrum using a vector network analyzer (VNA) (see Methods). Figure 6 shows the measured reflection spectra from Port 2 ($S_{22}$). It presents the amplitude of the reflection ($|S_{22}|$) in log scale (dB) in the two-dimensional map (frequency vs. voltage). The voltage in the *x*-axis is presented in decreasing order for easy comparison with the simulation data in Figure 5; the capacitance increases when the voltage applied to the varactor decreases.

The reflection spectra in Figure 6 show strong dips at the resonant mode frequency (yellow or blue in the map). Away from the resonant modes, the reflection remains very high (red in the map). Gradual changes in the overall features are observed as the metal-line angle varies from θ = 14° to 18°. At θ = 15° and 16°, a spectral overlap of the two resonant modes clearly appears, which is consistent with the eigenmode simulations in Figures 5 and S2. Two separate resonances with similar linewidths exist at zero bias. The two modes become closer in resonance frequency with increasing voltage (or decreasing capacitance). For θ = 16°, a spectral overlap of the originally separated resonances occurs at approximately 9–10 V. When the voltage increases further, the two modes remain nearly overlapped in the frequency spectrum. The overall behavior in the experiment agrees well with the eigenmode simulations.

To further confirm the EP behavior, the measured reflection amplitudes ($|S_{22}|$) at θ = 15° and 16° were directly fitted to the following dual Lorentzian function:

$$|S_{22}| = \left| \frac{a_1}{i(f-a_2)+a_3} + \frac{b_1}{i(f-b_2)+b_3} + c \right|. \tag{4}$$

The resonance frequency ($a_2$, $b_2$) and damping rate ($a_3$, $b_3$) of the two modes (i.e., the real and imaginary parts of the complex resonance frequency) were determined via fitting. The FWHM linewidth in the reflection spectrum is twice the damping rate. Figures S6 and S7 in Supplementary Material provide more details on the fitting procedure. Figures 7A and 7B show the fitted real and imaginary parts for θ = 15°, respectively, while Figures 7C and 7D show the real and imaginary parts for θ = 16°, respectively. At θ = 15°, the coalescence points of the real and imaginary parts are slightly separated. However, at θ = 16°, both the real and imaginary parts nearly coalesce simultaneously around 9 V, indicating the existence of the EP. In our experiment (Figures 6C, 7C and 7D), the real part of the resonance frequency of the two modes splits into two below 9 V (i.e., strong coupling regime) while the imaginary parts remain very close. Above 9 V, the situation is reversed; the imaginary part splits into two. In this case, one mode has a larger damping rate (super-radiant mode), and the other mode has a smaller damping rate (sub-radiant mode). Therefore, our experiment (as well as simulations) manifests the general features of non-Hermitian systems and EPs very clearly [45].

When the reflection amplitude decreases to nearly zero, phase singularity can occur because the phase cannot be defined as a single value at the zero-reflection amplitude [46]. The reflection phase was also directly measured from the VNA measurements. Figure S8 shows the measured phase spectra around the EP at θ = 16° at several applied voltages. The phase profile changes abruptly when the voltage changes from 9 to 10 V, confirming the behavior of phase singularity.

The electrical addressing of EPs in plasmonic structures may be useful for realizing highly functional elements in compact device platforms [47, 48]. For example, strongly confined modes in spoof plasmon structures have been used in index sensing [49–51] and wireless sensor networks [52]. Because EPs can significantly enhance sensitivity under small perturbations [13], electrically addressable EPs in compact spoof plasmon structures may further enhance the sensitivity and enable ultrasensitive sensors in highly integrated platforms. In addition to sensitivity enhancement, EPs can also induce many unusual phenomena, including mode conversion via the encircling of an EP [5, 6, 53], dynamic slowing or stopping of electromagnetic waves [54], and topological control of light propagation [6, 14]. Electrically addressable EPs may enable novel chip-scale, highly functional devices by exploiting the exceptional features of EPs in integrated platforms.

## 3. Conclusions

We presented the electrical addressing of EPs in a compact single-resonator structure. In our design, a localized spoof plasmonic resonator was combined with an angled metal line, and the interaction between the plasmonic modes was electrically controlled using a varactor loaded in the gap between the plasmonic resonator and the surrounding metal plane. Continuous electrical tuning of the varactor capacitance facilitated the simultaneous coalescence of the real and imaginary parts of the eigenfrequency and thus allowed the direct addressing of EPs. The electrically addressable EPs studied in this work may enable highly functional and tunable elements in compact device platforms.

## 4. Methods

A compact plasmonic resonator was fabricated using a two-layer PCB. Plasmonic structures on the front side and a metal line on the back side were patterned using an 18 μm thick copper layer. Five samples with metal-line angles of $\theta = 14°–18°$ were prepared in increments of 1°. The PCB substrate is TLY-5 (Taconic; dielectric constant: 2.2, loss tangent: 0.0009, thickness: 0.254 mm). The outer radius ($r_1$) of the LSP structure on the front side is 18 mm, whereas the inner radius ($r_2$) is 10 mm. The angle of each metal wing is 4°, and the LSP structure has 46 wings. The metal line on the back side has a linewidth of 0.752 mm. For electrical measurements, a varactor (SMV2019-079LF, Skyworks) was soldered to the inner LSP structure and the outer surrounding metal plane. Additional copper wires were soldered to the center of the inner LSP structure and the corner of the outer metal plane to apply a bias. SubMiniature version A (SMA) connectors were used at ports. The signal line of the SMA connector was soldered to the metal line on the back side, while the concentric metal part of the SMA connector was soldered to the outer metal plane on the front side. The outer metal plane surrounding the LSP structure was used as the common ground for the varactor and SMA connectors.

The complex eigenfrequencies of the plasmonic structure for different metal-line angles and capacitances were obtained using COMSOL eigenvalue solver (RF module). Eigenfrequency/eigenmode simulations were performed to identify the resonant modes in the LSP structures. All simulations were conducted in 3-dimensional geometry inside a rectangular box that was designated as the scattering boundary condition. As for varactor modeling, the 2-dimensional geometry of the varactor (1.5 mm x 0.5 mm) was introduced instead of 3-dimensional geometry to prevent inadequate complex meshing. A lumped element was set to the varactor with the series RLC condition. The resistance of the varactor was set as 4.8 Ω while the lumped element inductance was set as 0.7 nH (following the specification sheet of the varactor used in the experiment). Then, a desired capacitance value was introduced. The varactor was placed in the gap between the spoof LSP structure and surrounding metal plane on the front side, which is the same as the experimental configuration. The copper patterns were modeled as perfect electric conductors in the simulations. The field profiles were obtained 1 mm away from the back side of the sample.

Reflection spectra were measured using a VNA (Keysight Technologies, N5242A PNA-X). All samples were mounted on a Styrofoam plate during the measurements (the refractive index of Styrofoam is close to 1 at microwave frequencies). A DC voltage of 0–20 V was applied to the varactor.

# References

[1] Ş. K. Özdemir, S. Rotter, F. Nori, and L. Yang, "Parity–time symmetry and exceptional points in photonics," *Nat. Mater.*, vol. 18, pp. 783–798, 2019.

[2] M. A. Miri, and A. Alù, "Exceptional points in optics and photonics," *Science*, vol. 363, p. eaar7709, 2019.

[3] M. Parto, Y. G. N. Liu, B. Bahari, M. Khajavikhan, and D. N. Christodoulides, "Non-Hermitian and topological photonics: optics at an exceptional point," *Nanophotonics*, vol. 10, pp. 403–423, 2021.

[4] H. Hodaei, A. U. Hassan, S. Wittek, et al., "Enhanced sensitivity at higher-order exceptional points," *Nature*, vol. 548, pp. 187–191, 2017.

[5] J. W. Yoon, Y. Choi, C. Hahn, et al., "Time-asymmetric loop around an exceptional point over the full optical communications band," *Nature*, vol. 562, pp. 86–90, 2018.

[6] Q. Song, M. Odeh, J. Zúñiga-Pérez, B. Kanté, and P. Genevet, "Plasmonic topological metasurface by encircling an exceptional point," *Science*, vol. 373, pp. 1133–1137, 2021.

[7] H. Cao, and J. Wiersig, "Dielectric microcavities: model systems for wave chaos and non-Hermitian physics," *Rev. Mod. Phys.*, vol. 87, p. 61, 2015.

[8] J. Zhu, Ş. K. Özdemir, L. He, and L. Yang, "Controlled manipulation of mode splitting in an optical microcavity by two Rayleigh scatterers," *Opt. Express*, vol. 18, pp. 23535–23543, 2010.

[9] J. Wiersig, "Structure of whispering-gallery modes in optical microdisks perturbed by nanoparticles," *Phys. Rev. A*, vol. 84, p. 063828, 2011.

[10] B. Peng, Ş. K. Özdemir, M. Liertzer, et al., "Chiral modes and directional lasing at exceptional points," *Proc. Natl. Acad. Sci. U.S.A.*, vol. 113, pp. 6845–6850, 2016.

[11] W. R. Sweeney, C. W. Hsu, S. Rotter, and A. D. Stone, "Perfectly absorbing exceptional points and chiral

absorbers," *Phys. Rev. Lett.*, vol. 122, p. 093901, 2019.

[12] W. Chen, Ş. Kaya Özdemir, G. Zhao, J. Wiersig, and L. Yang, "Exceptional points enhance sensing in an optical microcavity," *Nature*, vol. 548, pp. 192–196, 2017.

[13] J. Wiersig, "Review of exceptional point-based sensors," *Photonics Res.*, vol. 8, pp. 1457–1467, 2020.

[14] M. S. Ergoktas, S. Soleymani, N. Kakenov, et al., "Topological engineering of terahertz light using electrically tunable exceptional point singularities," *Science*, vol. 376, pp. 184–188, 2022.

[15] Y. Huang, Y. Shen, C. Min, and G. Veronis, "Switching of the direction of reflectionless light propagation at exceptional points in non-PT-symmetric structures using phase-change materials," Opt. Express, vol. 25, pp. 27283-27297, 2017.

[16] Y. Yao, M. A. Kats, P. Genevet, N. Yu, Y. Song, J. Kong, and F. Capasso, "Broad electrical tuning of graphene-loaded plasmonic antennas," Nano Lett., vol. 13, pp. 1257–1264, 2013.

[17] F.-Z. Shu, J.-N. Wang, R.-W. Peng, B. Xiong, R.-H. Fan, Y.-J. Gao, Y. Liu, D.-X. Qi, and M. Wang, "Electrically driven tunable broadband polarization states via active metasurfaces based on joule-heat-induced phase transition of vanadium dioxide," Laser Photonics Rev., vol. 15, p. 2100155, 2021.

[18] I. Kim, W.-S. Kim, K. Kim, M. A. Ansari, M. Q. Mehmood, T. Badloe, Y. Kim, J. Gwak, H. Lee, Y.-K. Kim, and J. Rho, "Holographic metasurface gas sensors for instantaneous visual alarms," Sci. Adv., vol. 7, p. eabe9943, 2021.

[19] J.-N. Wang, B. Xiong, R.-W. Peng, C.-Y. Li, B.-Q. Hou, C.-W. Chen, Y. Liu, and M. Wang, "Flexible phase change materials for electrically-tuned active absorbers," Small, vol. 17, p. 2101282, 2021.

[20] J. Pendry, L. Martin-Moreno, and F. Garcia-Vidal, "Mimicking surface plasmons with structured surfaces," *Science*, vol. 305, pp. 847–848, 2004.

[21] F. Garcia-Vidal, L. Martin-Moreno, and J. Pendry, "Surfaces with holes in them: new plasmonic metamaterials," *J. Opt. A: Pure Appl. Opt.*, vol. 7, p. S97, 2005.

[22] P. A. Huidobro, A. I. Fernández-Domínguez, J. B. Pendry, L. Martín-Moreno, and F. J. García-Vidal, *Spoof surface plasmon metamaterials*, Cambridge, Cambridge University Press, 2018.

[23] X. Shen, T. J. Cui, D. Martin-Cano, and F. J. Garcia-Vidal, "Conformal surface plasmons propagating on ultrathin and flexible films," *Proc. Natl. Acad. Sci. U.S.A.*, vol. 110, pp. 40–45, 2013.

[24] B. J. Yang, Y. J. Zhou, and Q. X. Xiao, "Spoof localized surface plasmons in corrugated ring structures excited by microstrip line," *Opt. Express*, vol. 23, pp. 21434–21442, 2015.

[25] F. J. Garcia-Vidal, A. I. Fernández-Domínguez, L. Martin-Moreno, H. C. Zhang, W. Tang, R. Peng, and T. J. Cui, "Spoof surface plasmon photonics," *Rev. Mod. Phys.*, vol. 94, p. 025004 (2022).

[26] S. A. Maier, *Plasmonics: Fundamentals and Applications*, New York, Springer, 2007.

[27] J. A. Schuller, E. S. Barnard, W. Cai, Y. C. Jun, J. S. White, and M. L. Brongersma, "Plasmonics for extreme light concentration and manipulation," *Nat. Mater.*, vol. 9, pp. 193–204, 2010.

[28] A. Pors, E. Moreno, L. Martin-Moreno, J. B. Pendry, and F. J. Garcia-Vidal, "Localized spoof plasmons arise while texturing closed surfaces," *Phys. Rev. Lett.*, vol. 108, p. 223905, 2012.

[29] X. Shen, and T. J. Cui, "Ultrathin plasmonic metamaterial for spoof localized surface plasmons," *Laser Photonics Rev.*, vol. 8, pp. 137–145, 2014.

[30] Z. Liao, A. I. Fernández-Domínguez, J. Zhang, S. A. Maier, T. J. Cui, and Y. Luo, "Homogenous metamaterial

description of localized spoof plasmons in spiral geometries," *ACS Photonics*, vol. 3, pp. 1768–1775, 2016.

[31] F. Gao, Z. Gao, Y. Zhang, X. Shi, Z. Yang, and B. Zhang, "Vertical transport of subwavelength localized surface electromagnetic modes," *Laser Photonics Rev.*, vol. 9, pp. 571–576, 2015.

[32] F. Gao, Z. Gao, Y. Luo, and B. Zhang, "Invisibility dips of near-field energy transport in a spoof plasmonic metadimer," *Adv. Funct. Mater.*, vol. 26, pp. 8307–8312, 2016.

[33] J. Zhang, Z. Liao, Y. Luo, X. Shen, S. A. Maier, and T. J. Cui, "Spoof plasmon hybridization," *Laser Photonics Rev.*, vol. 11, p. 1600191, 2017.

[34] P. Qin, Y. Yang, M. Y. Musa, et al., "Toroidal localized spoof plasmons on compact metadisks," *Adv. Sci.*, vol. 5, p. 1700487, 2018.

[35] Z. Liao, G. Q. Luo, B. G. Cai, B. C. Pan, and W. H. Cao, "Subwavelength negative-index waveguiding enabled by coupled spoof magnetic localized surface plasmons," *Photonics Res.*, vol. 7, pp. 274–282, 2019.

[36] B. Sun, and Y. Yu, "Double toroidal spoof localized surface plasmon resonance excited by two types of coupling mechanisms," *Opt. Lett.*, vol. 44, pp. 1444-1447, 2019.

[37] Q. Zhou, Y. Fu, J. Liu, et al., "Plasmonic bound states in the continuum in compact nanostructures," *Adv. Opt. Mater.*, vo. 10, p. 2201590, 2022.

[38] H. C. Zhang, P. H. He, X. Gao, W. X. Tang, and T. J. Cui, "Pass-band reconfigurable spoof surface plasmon polaritons," *J. Phys. Condens. Matter.*, vol. 30, p. 134004, 2018.

[39] Q. Le Zhang and C. H. Chan, "Spoof surface plasmon polariton filter with reconfigurable dual and non-linear notched characteristics," *IEEE Trans. Circuits Syst. II Express Briefs*, vol. 68, pp. 2815–2819, 2021.

[40] X. Gao, H. C. Zhang, L. W. Wu, et al., "Programmable multifunctional device based on spoof surface plasmon polaritons," *IEEE Trans. Antennas Propag.*, vol. 68, pp. 3770–3779, 2020.

[41] Z. Liao, Y. Luo, A. I. Fernández-Domínguez, X. Shen, S. A. Maier, and T. J. Cui, "High-order localized spoof surface plasmon resonances and experimental verifications," Sci. Rep., vol. 5, p. 9590, 2015.

[42] X. Zhang, D. Bao, J. F. Liu, and T. J. Cui, "Wide-bandpass filtering due to multipole resonances of spoof localized surface plasmons," Ann. Phys. (Berlin), vol. 530, p. 1800207, 2018.

[43] Y. Lim, S.-C. An, H. Y. Jeong, T. H.-Y. Nguyen, G. Byun, and Y. C. Jun, "Multipole resonance and Vernier effect in compact and flexible plasmonic structures," *Sci. Rep.*, vol. 11, p. 22817, 2021.

[44] Y. Moritake and M. Notomi, "Switchable unidirectional radiation from Huygens dipole formed at an exceptional point in non-Hermitian plasmonic systems," ACS Photon., vol. 10, pp. 667−672, 2023.

[45] K. Zhang, Y. Xu, T.-Y. Chen, H. Jing, W.-B. Shi, B. Xiong, R.-W. Peng, and M. Wang, "Multimode photon-exciton coupling in an organic-dye-attached photonic quasicrystal," Opt. Lett., vol. 41, p. 5740, 2016.

[46] H. Y. Jeong, Y. Lim, S. C. An, T. H. Y. Nguyen, G. Byun, and Y. C. Jun, "Tunable resonance and phase vortices in kirigami Fano-resonant metamaterials," *Adv. Mater. Technol.*, vol. 5, p. 2000234, 2020.

[47] Z. Liao, X. Shen, B. C. Pan, J. Zhao, Y. Luo, and T. J. Cui, "Combined system for efficient excitation and capture of LSP resonances and flexible control of SPP transmissions," *ACS Photonics*, vol. 2, pp. 738–743, 2015.

[48] X. Gao, Z. Gu, Q. Ma, W. Y. Cui, T. J. Cui, and C. H. Chan, "Reprogrammable spoof plasmonic modulator," *Adv. Funct. Mater.*, DOI: 10.1002/adfm.202212328.

[49] J. Cai, Y. J. Zhou, Y. Zhang, and Q. Y. Li, "Gain-assisted ultra-high-Q spoof plasmonic resonator for the sensing of polar liquids," *Opt. Express*, vol. 26, pp. 25460–25470, 2018.

[50] Y. J. Zhou, Q. Y. Li, H. Z. Zhao, and T. J. Cui, "Gain-assisted active spoof plasmonic Fano resonance for high-resolution sensing of glucose aqueous solutions," *Adv. Mater. Technol.*, vol. 5, p. 1900767, 2020.

[51] X. Zhang, W. Y. Cui, Y. Lei, X. Zheng, J. Zhang, and T. J. Cui, "Spoof localized surface plasmons for sensing applications," *Adv. Mater. Technol.*, vol. 6, p. 2000863, 2021.

[52] X. Tian, P. M. Lee, Y. J. Tan, et al., "Wireless body sensor networks based on metamaterial textiles," *Nat. Electron.*, vol. 2, pp. 243–251, 2019.

[53] J. Doppler, A. A. Mailybaev, J. Böhm, et al., "Dynamically encircling an exceptional point for asymmetric mode switching," *Nature*, vol. 537, pp. 76–79, 2016.

[54] T. Goldzak, A. A. Mailybaev, and N. Moiseyev, "Light stops at exceptional points," *Phys. Rev. Lett.*, vol. 120, p. 013901, 2018.

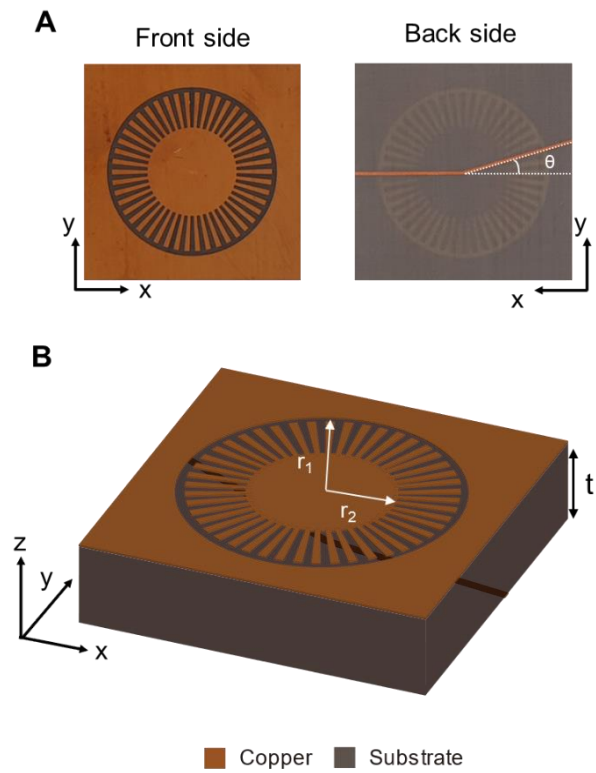

**Figure 1.** Localized spoof plasmon (LSP) structure used in this work. (A) Images of the front and back sides. The angle of the metal line on the back side is θ. (B) Schematic of the LSP structure ($r_1$ = 18 mm, $r_2$ = 10 mm, and t = 0.254 mm). The substrate appears to be semi-transparent in the schematic to help conceptual understanding. The thickness of the substrate in the schematic is pictured significantly larger than the real size for ease of recognition.

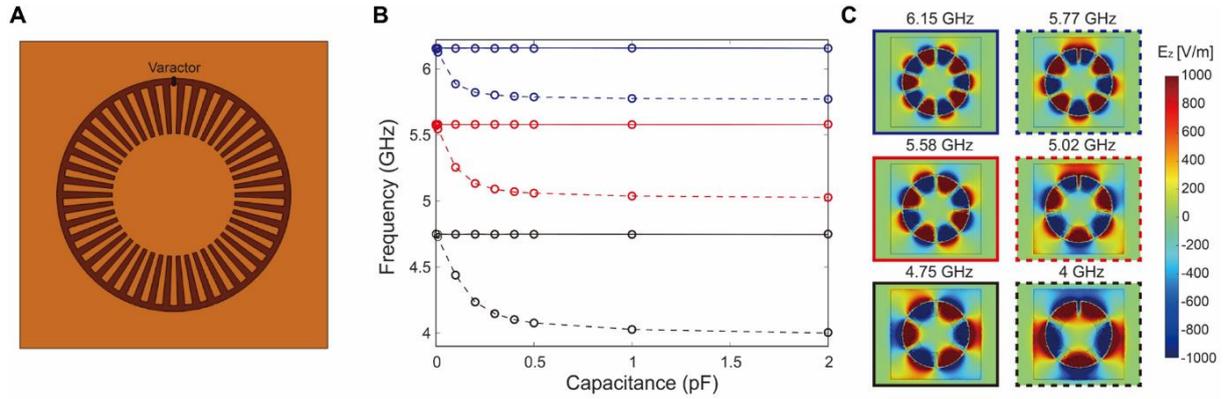

**Figure 2.** Mode perturbation by a varactor. (A) Schematic of the simulation geometry. A varactor-loaded spoof plasmon structure is patterned on the substrate without the metal line on the back side (i.e., simplified structure). The varactor is directly loaded on the wing of the spoof plasmon structure and the surrounding ground plane on the front side without any additional pads. (B) Initially at zero capacitance, there exist two degenerate modes for each resonance. As the capacitance increases, one of them gradually redshifts (dashed line) while the other mode remains unaffected (solid line). (C) Eigenmode field profiles ($E_z$) at a capacitance of 2 pF. The profiles in the left (solid box) and right (dashed box) columns correspond to solid and dashed lines in (B), respectively. When the capacitor is located at non-zero field positions, gradual redshift occurs with increasing capacitance.

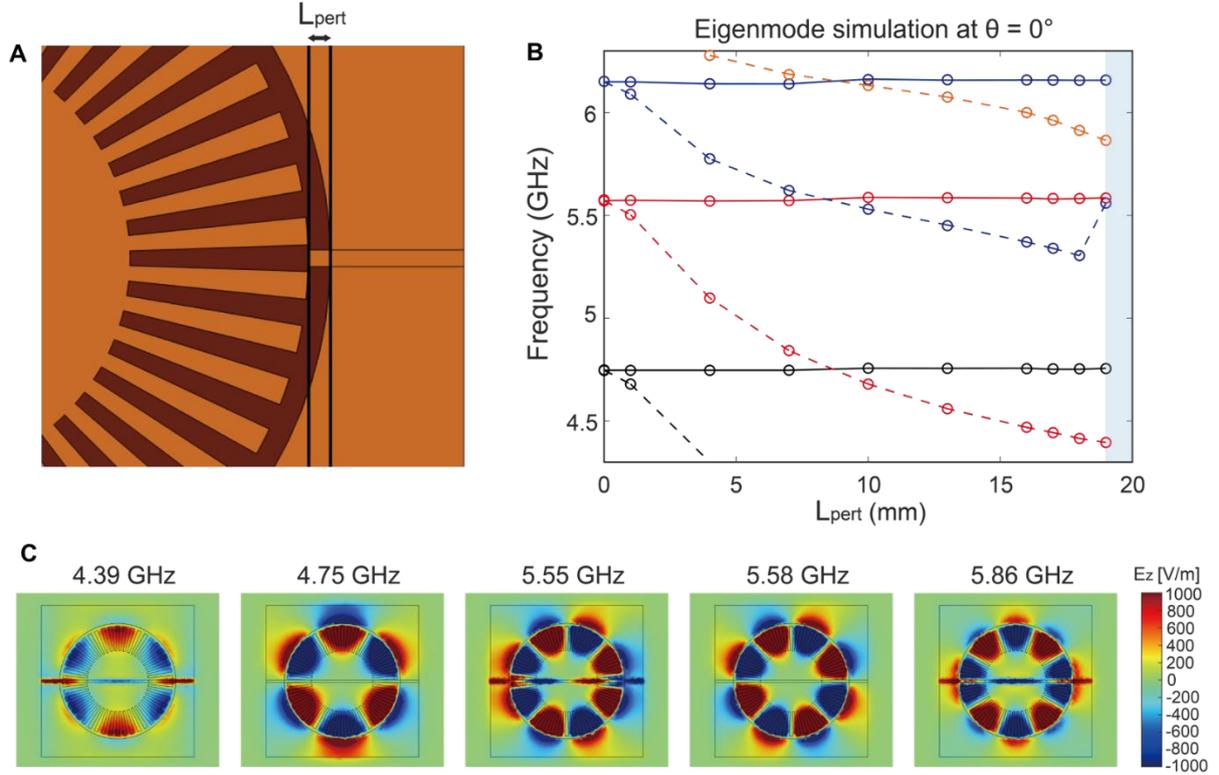

**Figure 3.** Mode perturbation by a metal line on the backside. (A) Schematic of the simulation geometry (having no varactor). The length of the two metal-line segments ($L_{pert}$) is gradually increasing until they touch ($L_{pert}$ = 19 mm) and form a continuous straight line ($\theta = 0°$). (B) Frequencies of eigenmodes with increasing $L_{pert}$, which are obtained from numerical simulations. Two degenerate modes exist for each resonance at zero segment length. With increasing segment length, the resonant frequency of one mode gradually redshifts while the resonant frequency of the other mode remains almost constant. (C) Eigenmode field profiles ($E_z$) when the two metal lines touched. When the metal line is located at the node position (e.g., 4.75 GHz, 5.58 GHz), the resonance frequency remains nearly constant regardless of the segment length $L_{pert}$. However, when the metal line lies at non-zero field positions, the eigenmode frequency gradually shifts with the segment length.

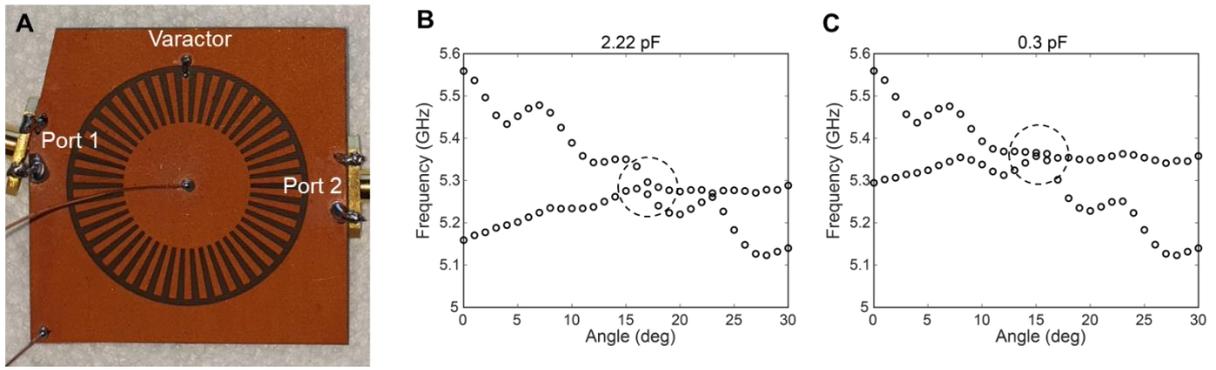

**Figure 4.** (A) Image of the real sample (with an angled metal line on the back side). A varactor was placed in the gap between the spoof LSP structure and the surrounding metal plane on the front side. Copper wires were soldered at the center of the spoof plasmon structure and at the corner of the outer metal plane to apply a bias across the gap. A cutting corner is introduced at Port 1 for the angled metal line. The angled metal line is cut normal to the cutting corner to induce the same geometry as the other port. (B) and (C) Eigenmode simulations with varying metal-line angles (from θ = 0° to 30°) for given capacitances of 2.22 pF and 0.3 pF, respectively. Two eigenmodes exist in the frequency range of 5 ~ 5.5 GHz, and they have interactions in the region of dashed circles.

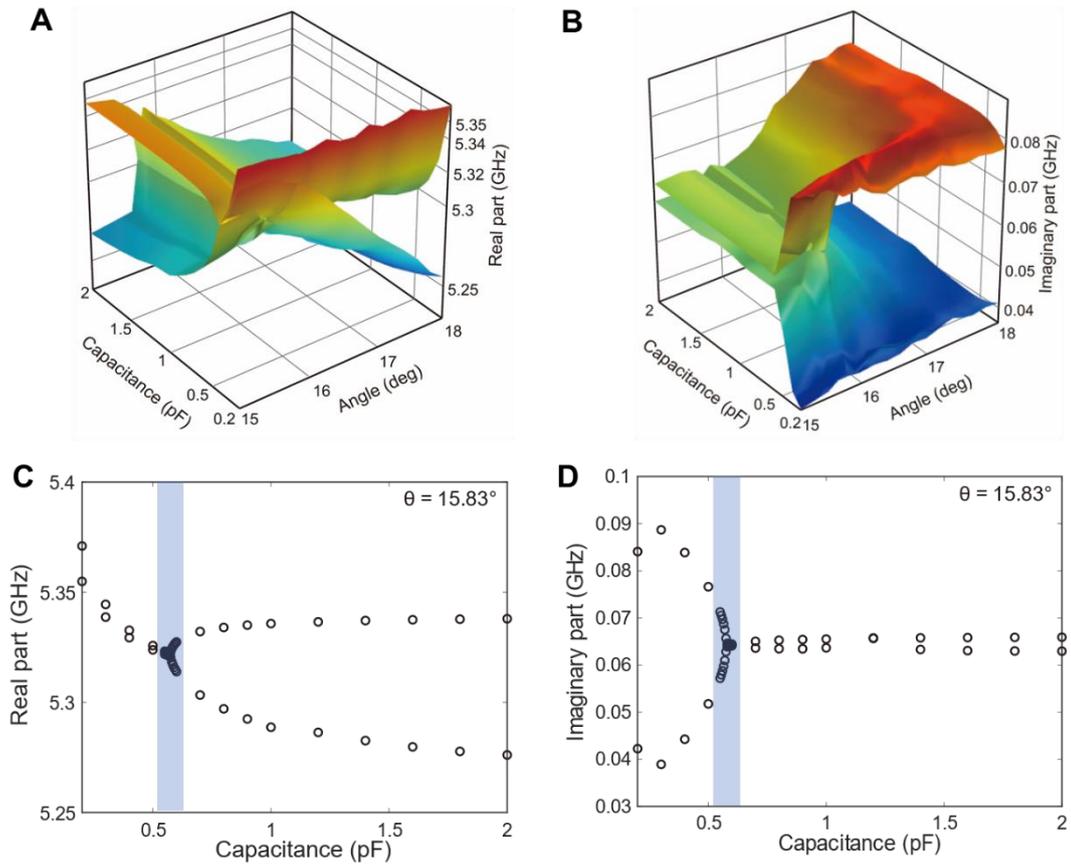

**Figure 5.** Simulated complex eigenfrequencies in parameter space (capacitance and metal-line angle). (A) and (B) Real and Imaginary parts of the complex eigenfrequency, respectively. They show the interactions of the two eigenmodes and a nontrivial topology in parameter space. (C) and (D) Real and Imaginary parts of the complex eigenfrequency at $\theta = 15.83°$, respectively. The two eigenmodes nearly coalesce in the shaded region in both the real and imaginary parts, manifesting the feature of the EP.

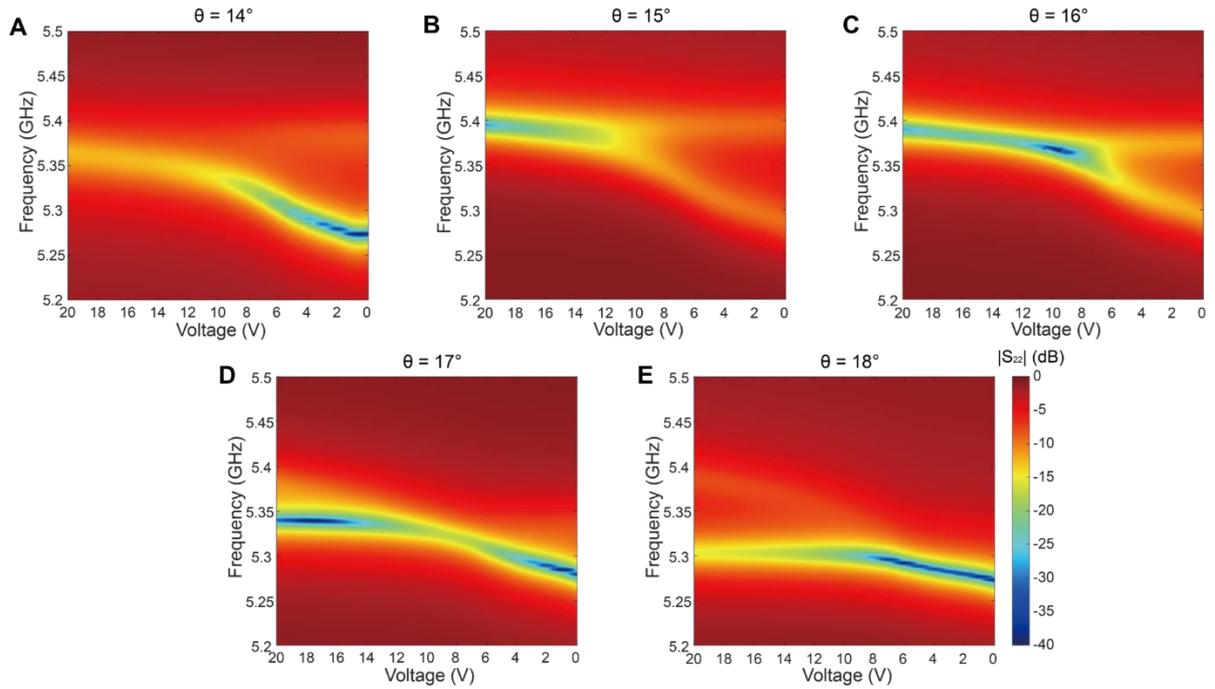

**Figure 6.** Experimental measurement data for different metal-line angles. Gradual changes are observed in the reflection spectrum when the metal-line angle θ changes. (A) – (E) Measured reflection amplitude ($|S_{22}|$) in logscale (dB) for θ = 14°, 15°, 16°, 17°, and 18°, respectively. The bias voltage applied to the varactor is tuned from 0 to 20 V. Note that the voltage in the *x*-axis is presented in decreasing order for easy comparison with the simulation data. The reflection spectrum shows a strong dip at the resonant mode frequency. At θ = 15° and 16°, the spectral separation and overlap of the two resonant modes are clearly visible.

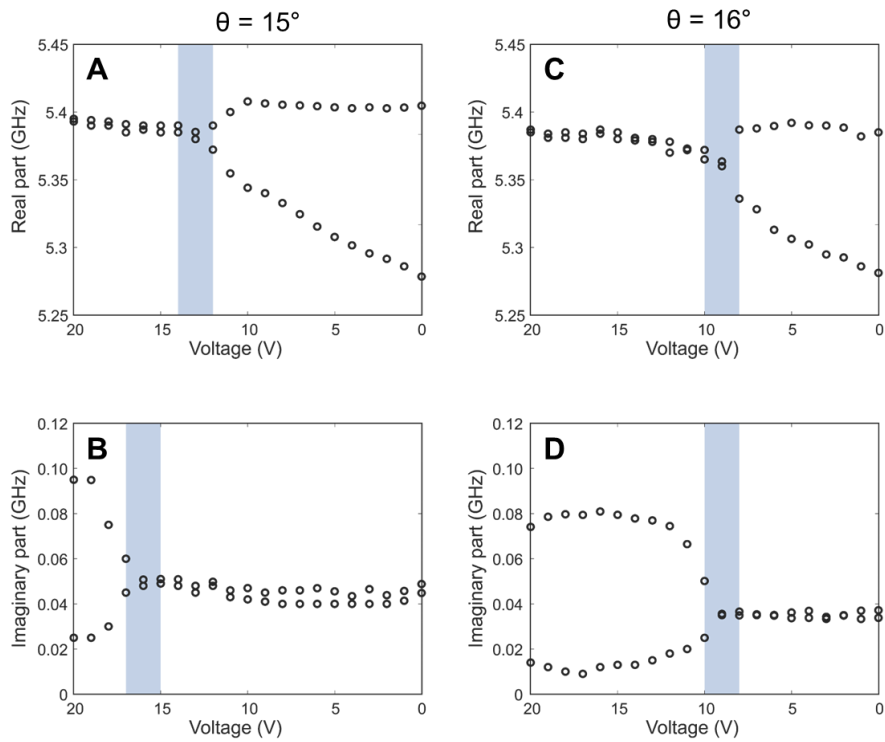

**Figure 7.** Fitting of the experimental reflection spectra to a dual Lorentzian function. (A) and (B) Real and imaginary parts of the complex resonance frequency for θ = 15°, respectively. (C) and (D) Real and imaginary parts for θ = 16°, respectively. The shaded parts indicate the voltage range where the two modes nearly coincide. At θ = 16°, both real and imaginary parts nearly coalesce around 9 V, while the coalescence points for the real and imaginary parts are more separated for θ = 15°.

# Supplementary Material

# Electrical addressing of exceptional points in compact plasmonic structures

- **Figure S1:** Snapshot of the time-domain simulation
- **Figure S2:** Simulated complex eigenfrequencies for different metal-line angles.
- **Figure S3:** Simulated eigenmode profile away from the EP condition
- **Figure S4:** Simulated eigenmode profile near the EP condition
- **Figure S5:** Mode-splitting simulations near the EP
- **Figure S6:** Experimental reflection spectra together with fitting curves
- **Figure S7:** Simulated reflection spectrum together with a fitting curve
- **Figure S8:** Experimentally measured phase spectra

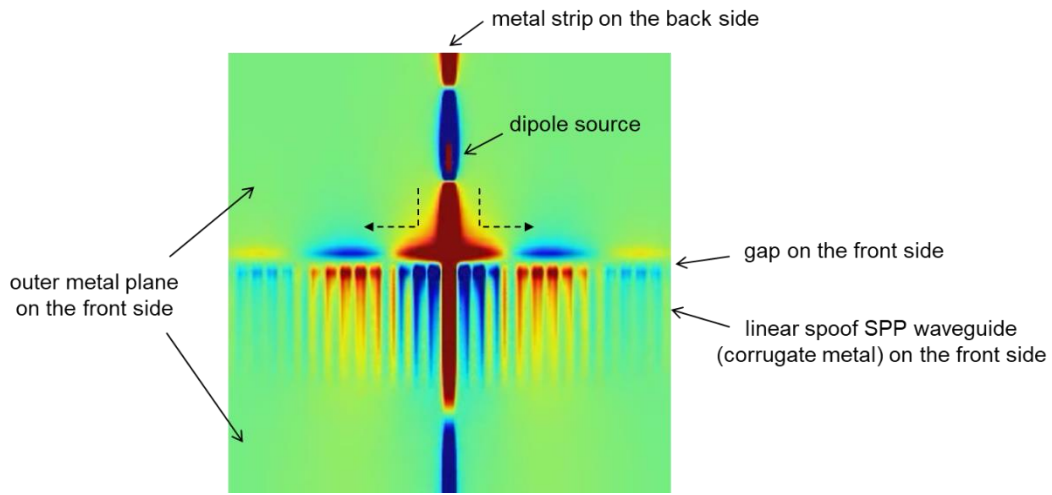

**Figure S1.** Snapshot of the time-domain simulation (Lumerical FDTD). The metal line on the backside is aligned to pass the middle of a one-dimensional (straight) spoof plasmon waveguide. A propagating wave along the metal line is directly excited by a point dipole source. The metal line works as a splitter (see Supplementary Video V1). Incident waves are split into the two opposite directions of the spoof plasmon waveguide. In the case of our circular localized spoof plasmon (LSP) resonator, the split waves become clockwise and counterclockwise circulating plasmon waves. Their interactions form different modes depending on the metal-line angle.

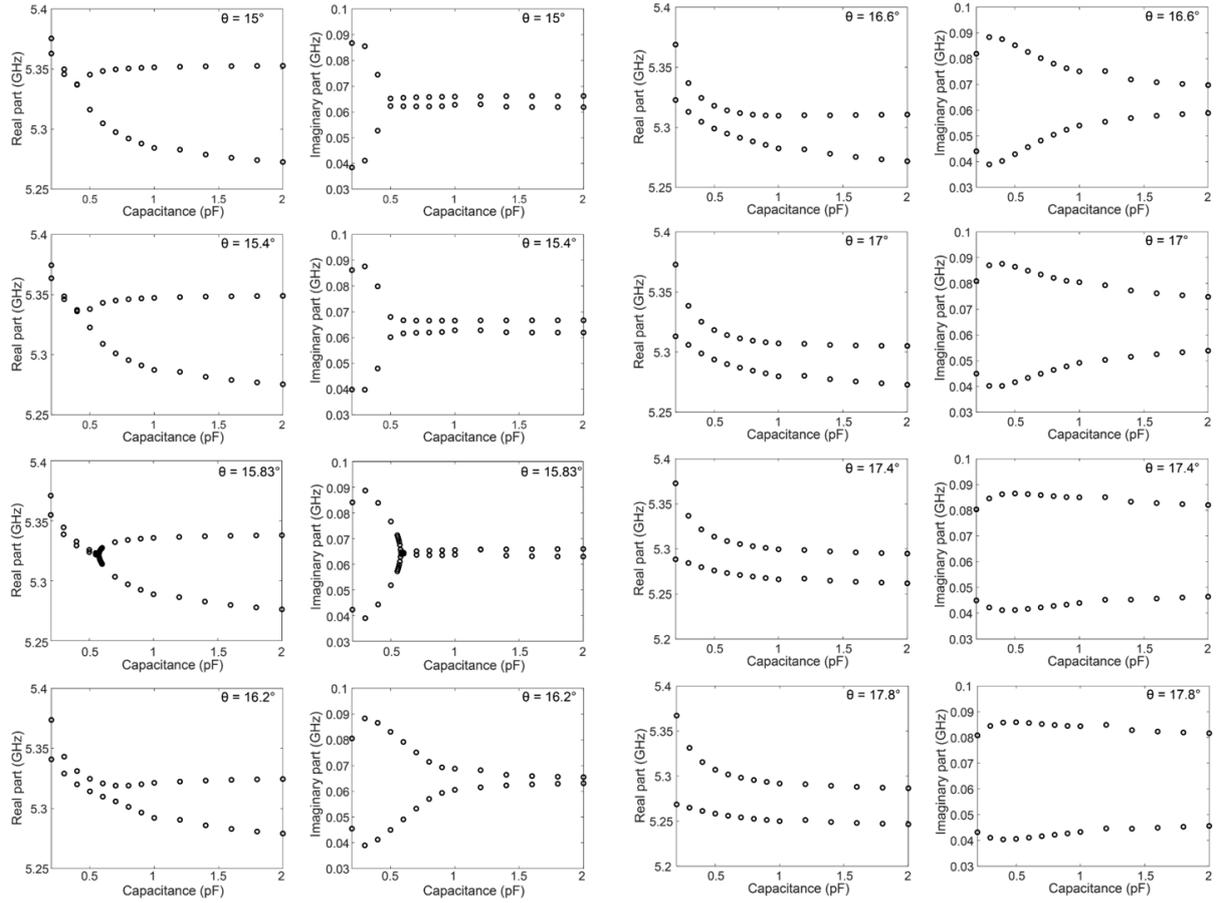

**Figure S2.** Simulated complex eigenfrequencies for different metal-line angles. Real and Imaginary parts of the complex eigenfrequencies are shown for θ = 15° ~ 17.8°. Between θ = 15° and 16°, both real and imaginary parts of the two eigenmodes become very close. The real and imaginary parts of the complex eigenfrequencies of the two eigenmodes nearly coalesce simultaneously at θ = 15.83°, corresponding to an exceptional point (EP).

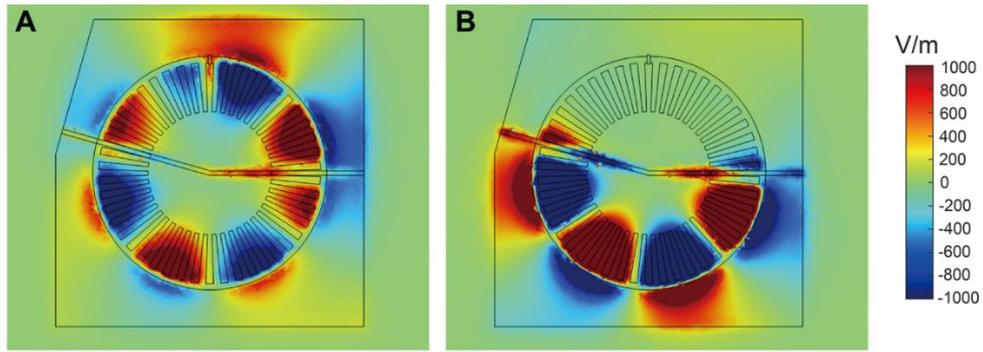

**Figure S3.** Simulated eigenmode profile ($E_z$) at $\theta = 15.83°$ and capacitance of 2 pF (away from the EP condition). (A) and (B) show the field profiles at 5.27 GHz and 5.34 GHz, respectively. A cutting corner was introduced at the upper-left corner for the angled metal line.

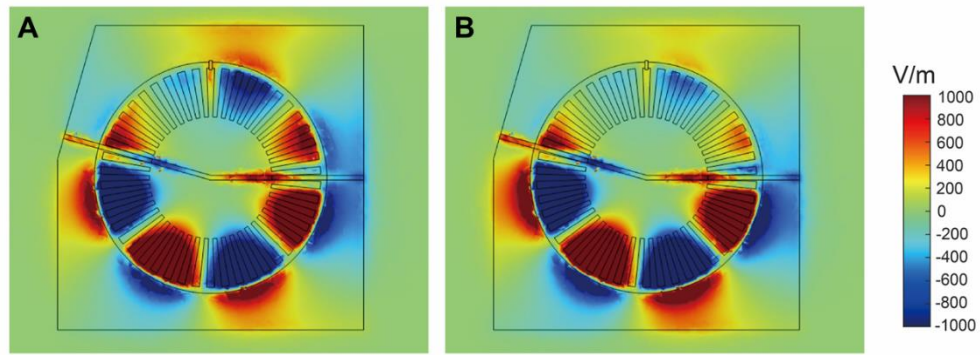

**Figure S4.** Simulated eigenmode profile ($E_z$) at $\theta = 15.83°$ and capacitance of 0.585 pF (near the EP condition). (A) and (B) are the field profiles at 5.32 GHz. The field profiles of the two eigenmodes become similar.

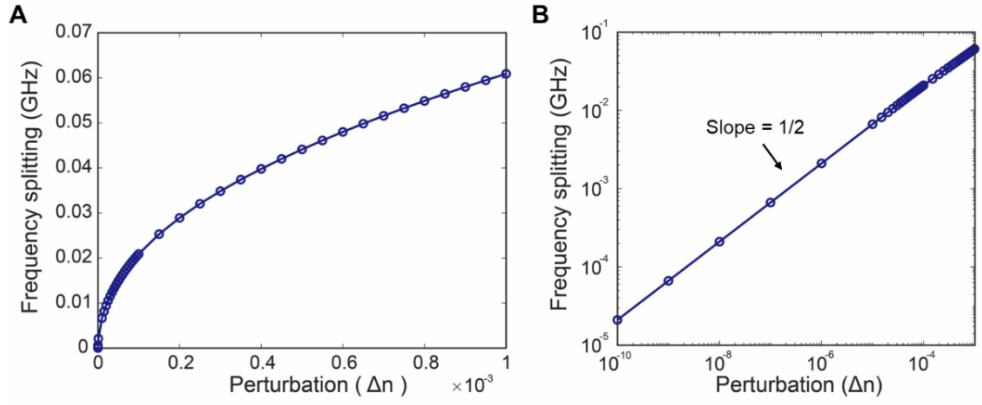

**Figure S5.** Mode-splitting simulations near the EP in our spoof plasmon structure. In the simulation, the refractive index of the surrounding medium is gradually increased (n = 1 + Δn), and changes in the mode splitting are measured. Fitting of this mode splitting in the log-log plot perfectly matches a square-root curve (i.e., slope = 1/2) under small perturbations. This is clearly different from other mode-splitting mechanisms (diabolic points) which exhibit linear dependence under perturbations.

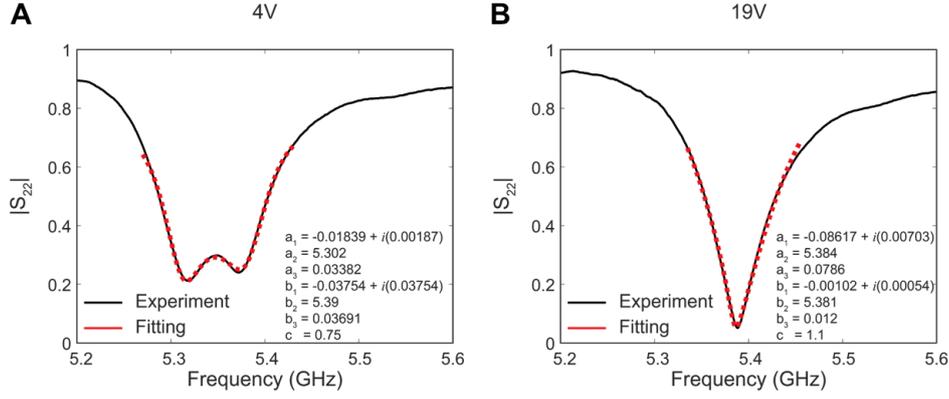

**Figure S6.** (A) and (B) Experimental reflection spectra at 4 V and 19 V, respectively, together with fitting curves (red dotted lines). The metal-line angle is θ = 16° in both cases. Fitting parameters are also indicated in each case. The number of fitting parameters was minimized in the initial fitting using proper simplifications. Then, fitting was repeated for all parameters to fine tune and optimize the fitting parameters.

(i) Figure S6A shows the experimental reflection spectrum at 4 V, where the reflection spectrum shows two, separate resonance dips. In this case, initial fitting was conducted separately for individual resonances. Then, the whole spectrum was fitted again for fine tuning.

(ii) Figure S6B shows the experimental reflection spectrum at 19 V (away from the EP condition), where the reflection spectrum shows a single (nearly overlapped) resonance dip. In this case, initial fitting was conducted using reasonable estimations for the resonance frequencies ($a_2 \approx b_2$) and the background level $c$. Then, fitting was repeated to optimize the whole fitting parameters.

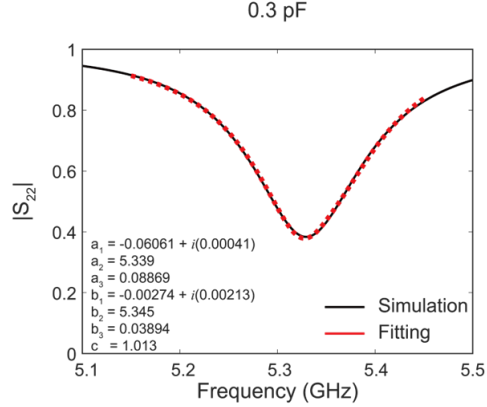

**Figure S7.** Simulated reflection spectrum at a capacitance of 0.3 pF together with a fitting curve (red dotted line) (θ = 15.83°). Fitting parameters are also indicated.

To confirm the experimental data fitting further, fitting is also conducted using the simulated reflection spectra. In this case, the real and imaginary parts of the complex resonance frequencies ($a_2$, $a_3$, $b_2$, $b_3$) are obtained from separate simulations (eigenfrequency simulations). Then, Eq. (4) was fitted to the simulated reflection spectrum to determine other parameters. Figure S7 shows an exemplary case at a capacitance of 0.3 pF (away from the EP condition) in Figure 5 (i.e., nearly overlapped resonance at θ = 15.83°). It is found that an overall trend in the fitting parameters of the simulation is very similar to that in the experimental data fitting. For example, for the nearly overlapped resonance, the high-Q mode ($|b_3| < |a_3|$) has a much smaller resonance amplitude ($|b_1/b_3| < |a_1/a_3|$) in both experiment and simulation (Figures S6B and S7). (However, we note that the experimental reflection spectrum has a sharper resonance and smaller $b_3$ than the simulated spectrum)

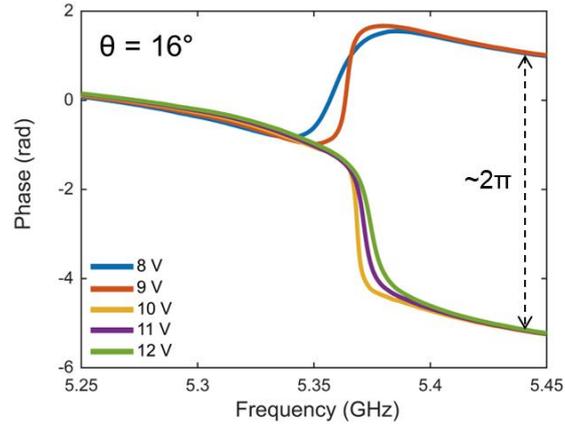

**Figure S8.** Experimentally measured phase spectra for several different voltages (θ = 16°). The phase spectra were also measured at Port 2. When the voltage changes from 9 to 10 V, an abrupt change in the phase spectrum occurs owing to the phase singularity.